\RequirePackage{fix-cm}
\documentclass{mysvjour3}                     % onecolumn (standard format)
\smartqed  % flush right qed marks, e.g. at end of proof
\usepackage{graphicx}
\usepackage{amsmath}
\usepackage{amssymb}
\usepackage{subfigure}
\usepackage{url}

\DeclareMathOperator{\tr}{tr}
\DeclareMathOperator{\COV}{COV}
\DeclareMathOperator{\Diag}{Diag}
\begin{document}

\title{Visualizing dimensionality reduction of systems biology data
%Statistical dimensionality reduction and visualization of systems biology data
\thanks{M.~Huber was supported by the Deutsche Forschungsgemeinschaft (DFG) via a Heisenberg grant (Hu954/4) and a Heinz Maier-Leibnitz Prize grant (Hu954/5). A.C.~Polatkan and A.~Pritzkau were supported by the DFG Priority Program 1335 ``Scalable Visual Analytics''.}
}


\author{Andreas Lehrmann \and Michael Huber \and Aydin C. Polatkan \and Albert Pritzkau \and Kay Nieselt}


\institute{A.~Lehrmann \at
              Center for Bioinformatics T\"ubingen,
              University of T\"ubingen\\
              Sand 14, 72076 T\"ubingen\\
              Tel.: ++49-7071-2970443,
              Fax: ++49-7071-295147\\
              \email{lehrmann@informatik.uni-tuebingen.de}           %  \\
           \and
          M.~Huber \at
              Wilhelm Schickard Institute for Computer Science,
              University of T\"ubingen\\
              Sand 13, 72076 T\"ubingen\\
              Tel.: ++49-7071-2977173, Fax: ++49-7071-295061\\	
              \email{michael.huber@uni-tuebingen.de}           %  \\
             \and
            A.C.~Polatkan \at
              Center for Bioinformatics T\"ubingen,
              University of T\"ubingen\\
              Sand 14, 72076 T\"ubingen\\
              Tel.: ++49-7071-2970443,
              Fax: ++49-7071-295147\\
              \email{polatkan@informatik.uni-tuebingen.de}                    
               \and
            A.~Pritzkau \at
              Bild- und Signalverarbeitung,
              University of Leipzig\\
              Johannisgasse 26, 04009 Leipzig\\
              \email{albert.pritzkau@medizin.uni-leipzig.de}          
                           \and
            K.~Nieselt \at
              Center for Bioinformatics T\"ubingen,
              University of T\"ubingen\\
              Sand 14, 72076 T\"ubingen\\
              Tel.: ++49-7071-2970443,
              Fax: ++49-7071-295147\\
              \email{kay.nieselt@uni-tuebingen.de}           %  \\              
}

\date{Received: September 8, 2011 / Accepted: April 23, 2012}

\maketitle

\begin{abstract}
One of the challenges in analyzing high-dimensional expression data is the detection of important biological signals. A common approach is to apply a dimension reduction method, such as principal component analysis. Typically, after application of such a method the data is projected and visualized in the new coordinate system, using scatter plots or profile plots. These methods provide good results if the data have certain properties which become visible in the new coordinate system and which were hard to detect in the original coordinate system. Often however, the application of only one method does not suffice to capture all important signals. Therefore several methods addressing different aspects of the data need to be applied. We have developed a framework for linear and non-linear dimension reduction methods within our visual analytics pipeline SpRay. This includes measures that assist the interpretation of the factorization result. Different visualizations of these measures can be combined with functional annotations that support the interpretation of the results. We show an application to high-resolution time series microarray data in the antibiotic-producing organism {\em Streptomyces coelicolor} as well as to microarray data measuring expression of cells with normal karyotype and cells with trisomies of human chromosomes 13 and 21.

\end{abstract}

\section{Introduction}
\label{intro}
In biology as well as in many other scientific areas, high dimensional data sets arise in a natural way. In the case of gene expression data, this is mostly a consequence of the widespread utilization of DNA microarrays \cite{Schena1995,Lockhart1996} and the development of high throughput sequencing methods \cite{Shendure2008}, especially during the course of the last decade. Both techniques allow the parallel analysis of thousands of genes and have thus led to an unprecedented flood of data. In order to find important signals in the large amount of data, a common approach is to combine feature extraction with subsequent dimensionality reduction. With a reduced dimensionality comes the advantage that the essential signals can easily be visualized either with the help of a parallel coordinates plot (PCP) as first introduced by Inselberg \cite{Inselberg1985,Inselberg2009} or, after a reduction to two or three dimensions, in a conventional two or three dimensional Cartesian coordinate system.

Among the vast variety of methods that are able to extract interesting signals out of large data sets are linear methods such as principal component analysis (PCA) \cite{Pearson1901,Hotelling1933,Joliffe2002} and independent component analysis (ICA) \cite{Hyvaerinen1997,Hyvaerinen1999,Hyvaerinen2001} as well as promising nonlinear procedures like the locally linear embedding (LLE) algorithm \cite{Roweis2000}. Although it is still a de facto standard, in our experience principal component analysis alone does not suffice to extract all important signals. Instead, it needs to be complemented with other methods in order to get the full picture \cite{Mannfolk2010}. For this reason, we have implemented several linear and nonlinear feature extraction methods including the ones mentioned above within our interactive and extendable visualization framework SpRay \cite{Dietzsch2009} that combines visual exploration with statistics-driven data analysis. 

After visualizing the low dimensional representation of an originally high dimensional data set, a common problem concerns the interpretation of the obtained result. In this work, we put special focus on high dimensional data sets that are structured in the sense that the succession of features or experiments is not arbitrary but allows for a meaningful ordering. 
Examples are time series gene expression or similar systems biology experiments.
If so, there is often substantial correlation between two adjacent features or experiments that transforms to a similar location in $n$-dimensional Euclidean space. Visualization methods that reflect this inherent structure in order to support the user during the process of dimensionality reduction are therefore of great use. Here, we introduce two new visualization methods that we call `loading maps' and `$l$-neighborhood evolution matrices'. They assist the interpretation in case of the linear methods and give further insights into the structure of high dimensional data sets which help during parameter selection in the nonlinear case.

In an application study, we demonstrate the usefulness and potential of these two visualizations by applying them to a high resolution time series experiment conducted on the bacterium \emph{Streptomyces coelicolor} \cite{Nieselt2010}. We show that loading maps are a helpful tool for the identification of principal or independent components that encapsulate signals from genomic clusters with common functional annotation. Furthermore, we demonstrate the ability of $l$-neighborhood evolution matrices to detect `time clusters' and significant biological events and to expose weak areas in strongly connected neighborhood graphs. The latter are an essential step not only in LLE but also in many other manifold learning algorithms (e.g., Isomap \cite{Tenenbaum2000}, Local Tangent Space Alignment (LTSA) \cite{Zhang2004} and Maximum Variance Unfolding (MVU) \cite{Weinberger2006}).

In a second application, we conduct a principal component analysis on the samples of a microarray data set measuring expression of cells with normal karyotype and cells with trisomies. Here, the primary interest is to identify prevalent expression profiles among samples, regardless of individual genes’ expression patterns. We show that the loading map again provides a helpful overview for the comparison of the principal components allowing, in this case, to quickly identify significant samples.

Over time, a lot of effort has been put into solving the problems mentioned so far: A classic method to facilitate the interpretation of principal components is the varimax criterion developed by Kaiser \cite{Kaiser1958}. For a visual impression of the dependency between the variables and the first two principal components, the circle of correlation \cite{Abdi2010} can be used. A novel approach using a so-called projection score to determine a suitable subset of variables that is easier to interpret was reported by Fontes and Soneson \cite{Fontes2011}. As far as LLE parameter selection is concerned, all available approaches are, to the best of our knowledge, of a purely computational nature. They range from hierarchical methods using residual variances \cite{Kouropteva2002} and methods using Spearman's rho \cite{Karbauskaite2007} to measures that assess the preservation of local geometry \cite{Valencia-Aguirre2009}. Our visual depiction of the neighborhood graph complements these mathematical methods in an excellent way. An interactive visualization environment was presented by Jeong \emph{et al.} \cite{Jeong2009}. However, their framework is limited solely to the analysis of principal components and lacks the possibility of comparing results obtained by using different methods. The same is true for many programs that offer gene expression analysis, e.g., GeneSpring \cite{Agilent}, TM4 \cite{Saeed2006} and Mayday \cite{Battke2010}.

In Section 2, we describe our mathematical framework, review linear and nonlinear feature extraction methods. Section 3 introduces our visualization methods in detail. Section 4 gives a short overview of SpRay and in Section 5 we present our application studies. We close with a discussion and an outlook.

\section{Dimension Reduction Methods} \label{sec:1}
\subsection{Theoretical framework}
Throughout the text, we assume the data to be represented by a matrix
\begin{equation}
X=\left(x_{ij}\right)_{\substack{1\leq i\leq n\\1\leq j\leq p}} \in \mathbb{R}^{n \times p},
\end{equation}
in which the rows represent a set of features (e.g., expression values) and the columns represent a set of experiments (e.g., a time series), that is, $x_{ij}$ is the value of a measurement of the $i$-th feature in the $j$-th experiment. When analyzing such a data set in a statistical environment, there are two perspectives that arise naturally: either we consider the features as variables and the experiments as observations or vice versa. To avoid case distinction, we will always assume in the following sections that the variables are given by the features. Subsequently, dimensionality reduction is done by a mapping
\begin{equation}
\phi: \mathbb{R}^n\longrightarrow \mathbb{R}^{n'}, x_{\bullet j} \mapsto \phi\left(x_{\bullet j}\right)=:y_{\bullet j}
\end{equation}
that maps every column vector $x_{\bullet j} \in \mathbb{R}^n$ of $X$ to a new column vector $y_{\bullet j} \in \mathbb{R}^{n'}$, thereby defining a dimensionality reduced matrix
\begin{equation}
Y=\left(y_{ij}\right)_{\substack{1\leq i\leq n'\\1\leq j\leq p}} \in \mathbb{R}^{n' \times p}.
\end{equation}
If $\phi\left(x\right)=Ax$ for an $n'\times n$ matrix $A$ (i.e., $Y=AX$), the dimensionality reduction method is called \emph{linear}, and \emph{nonlinear} otherwise. Often, $\phi$ is based on a certain cost function that one wants to maximize or minimize.

It is sometimes useful to think of a column of a matrix as being a realization of an underlying random vector. Where appropriate, we will switch between random vectors and their realizations in the form of a matrix without explicitly mentioning it. The same applies to statistical measures and their estimates.

Next, we give a short outline of the basic concepts of linear dimensionality reduction and then turn to a nonlinear approach.

\subsection{Linear dimensionality reduction}
Linear dimensionality reduction usually starts with centering of the data matrix $X$, a procedure that is equivalent to moving the coordinate system to the center of mass of the $n$-dimensional point cloud given by the columns of $X$. The coordinates with respect to the new coordinate system are now given by
\begin{equation}
X_c:=X\cdot\left(I_p-\frac{1}{p}\left(1,\cdots,1\right)^T\left(1,\cdots,1\right)\right),
\end{equation}
with $\bullet^T$ being the transpose of a matrix or vector, and $I_p$ being the $p$-dimensional identity matrix.

As pointed out earlier, we look for a matrix $A$, so that $AX_c$ optimizes a method-specific cost function. We want to briefly review two popular cost functions that are based on the maximization of variance and independence and lead to a principal component analysis and an independent component analysis, respectively.

%%%%%%%

\paragraph{Principal Component Analysis}
Principal Component Analysis (PCA) \cite{Joliffe2002} is a linear coordinate transformation that rotates a (centered) coordinate system in such a way that the coordinates with respect to the $i$-th axis have the maximum possible variance and are uncorrelated to the coordinates with respect to all axes $j<i$. Explicitly, we calculate the eigenvalue decomposition of the covariance matrix of $X_c$,
\begin{equation}
\COV\left[X_c\right]:=\left(p-1\right)^{-1}X_cX_c^T=Q\Lambda Q^T,
\end{equation}
and obtain the new coordinates satisfying the requirements as 
\begin{equation}\label{PCA_transformation}
Y_c=AX_c=Q^TX_c.
\end{equation}

The matrix $Q$ is the $n\times n$ matrix of the eigenvectors of the covariance matrix of $X_c$.
All eigenvalue decompositions are assumed to be ordered in the sense that $\Lambda=\Diag\left(\lambda_1,\ldots,\lambda_n\right)$ with $\lambda_1\geq\ldots\geq\lambda_n$. The rows of $Y_c$ are called \emph{principal components} and for the new data matrix $Y_c$, we have $\COV\left[Y_c\right]=\Lambda$ and $\tr\left(\COV\left[Y_c\right]\right)=\tr\left(\COV\left[X_c\right]\right)$, i.e., the rows of $Y_c$ are uncorrelated, $\mathbb{V}\left[y_{i\bullet}\right]=\lambda_i$ and the total variance has not changed.

The matrix $Y_c$ is still of the same dimensionality as $X_c$. A common method to reduce dimensionality is to choose some value $\zeta \in \left[0,1\right]$ specifying the amount of variance one wishes to maintain, to set
\begin{equation}
n':=\min\left\{k\in\left\{1,\ldots,n\right\}\middle| \frac{\sum_{i=1}^k \lambda_i}{\tr\left(\COV\left(Y_c\right)\right)}\geq\zeta\right\}
\end{equation}
and to limit the columns of $Q$ in equation \eqref{PCA_transformation} to $n'$. This is based on the assumption that important signals contribute the most to the variance and are captured by the first few principal components. 
\paragraph{Independent Component Analysis}
While independent random variables are always uncorrelated, the converse is generally not true. It is therefore a natural question to ask whether it is possible not only to decorrelate the data but also to maximize independence. This is indeed possible in several ways and Independent Component Analysis (ICA) \cite{Hyvaerinen2001} comprises a broad field of techniques which aims precisely at decorrelation and maximization of independence of the data at the same time. The basic principle is as follows: We assume that the centered data matrix $X_c$ is the result of linear mixing of some underlying matrix $Y_c$ whose rows are mutually independent and not normally distributed. The goal is to reconstruct this matrix $Y_c$ by finding the inverse $D$ of the (unknown) mixing matrix $M$ and applying  
\begin{equation}
Y_c:=DX_c.
\end{equation}
The ICA-model is greatly simplified by a whitening step that turns $X_c \in \mathbb{R}^{n\times p}$ into a new matrix $Z \in \mathbb{R}^{n\times p}$ with $\COV\left[Z\right]=I_n$. This corresponds to decorrelation by PCA-preprocessing with subsequent scaling of the variances. If the eigenvalue decomposition $\COV\left[X_c\right]=Q\Lambda Q^T$ has full rank, then
\begin{equation}
Z:= \sqrt{\Lambda^{-1}}Q^TX_c=\sqrt{\Lambda^{-1}}Q^TMY_c
\end{equation}
is white. Note, that the dimensions of both matrices $D$ and $M$ are independent of the input data.
If $\COV\left[X_c\right]$ is rank-limited (as is the case whenever $n\geq p$), the rows of $Q^T$ and the rows/columns of $\Lambda$ need to be limited to the rank $rk\left(\COV\left[X_c\right]\right)$. This is equivalent to a PCA-dimensionality reduction with subsequent scaling of the variances. Even in situations where it is not necessary, dimensionality reduction at this point is highly recommended. It is easy to see that we are now looking for the rows of an \emph{orthogonal} demixing matrix $D'=\left(d_{ij}'\right)$. A popular class of algorithms named FastICA to determine the rows of $D'$ was introduced by Hyvaerinen \cite{Hyvaerinen1997,Hyvaerinen1999} and is based on the maximization of non-Gaussianity. To maximize non-Gaussianity of the reconstruction $y_{i\bullet}:=d_{i\bullet}'Z$, one can either maximize (approximations of) negentropy $J\left[d_{i\bullet}'Z\right]$ \cite{Hyvaerinen1997a} or the absolute value of the fourth cumulant $\left\vert\kappa_4\left(d_{i\bullet}'Z\right)\right\vert$. The maximization of these values are computed in an iterative process involving gradient ascent or a fixed-point algorithm.

\subsection{Nonlinear dimensionality reduction}
\paragraph{Locally Linear Embedding}
The Locally Linear Embedding (LLE) \cite{Roweis2000} algorithm is based on the intuition that, although the columns of $X$ are points in $n$-dimensional Euclidean space, they might in fact lie on a much lower dimensional submanifold of dimension $n'\ll n$. In this case, the data set can be represented in global, internal coordinates of the submanifold. To this end, every column vector $x_{\bullet j}$ of $X$ is linearly approximated as well as possibly by $\sum_{k=1}^p w_{jk}x_{\bullet k}$. Defining $W=\left(w_{jk}\right)_{1\leq j,k\leq p} \in \mathbb{R}^{p\times p}$ as the matrix of the weights, this approach leads to the minimization of the $W$-dependent total reconstruction error
\begin{equation}
K_1\left(W\right)=\sum_{j=1}^p\left\lVert x_{\bullet j}-\sum_{k=1}^p w_{jk}x_{\bullet k}\right\rVert^2.
\end{equation}
To ensure locality and translational invariance, we enforce $w_{jk}=0$ for all $x_{\bullet k} \not\in N\left(x_{\bullet j}\right)$ and $\sum_{k=1}^p w_{jk}=1$ for all $1\leq j \leq p$. Here, $N$ is an arbitrary function that assigns each column vector of $X$ a set of neighbors (e.g., based on their Euclidean distance). By definition, $x_{\bullet j}$ can not belong to its own neighborhood $N\left(x_{\bullet j}\right)$.
Note that the `size and form' of the neighborhood of the individual points constitute the only free parameter of the algorithm. Therefore, its correct choice is of fundamental importance and crucially influences the result. We will come back to this important point shortly.

Because the optimal weights $w_{j\bullet}$ are invariant under translations, rotations and rescalings of $x_{\bullet j}\cup N\left(x_{\bullet j}\right)$, they can, once found, be fixed, and in a second step, the cost function is minimized subject to the embedding, i.e., we now minimize
\begin{equation}
K_2\left(Y\right)=\sum_{j=1}^p\left\lVert y_{\bullet j}-\sum_{k=1}^p w_{jk}y_{\bullet k}\right\rVert^2.
\end{equation} 
This can be rewritten as
\begin{equation}
\sum_{i=1}^{n'} y_{i\bullet}\underbrace{\left(I_p-W\right)^T\left(I_p-W\right)}_{=:M}y_{i\bullet}^T = \tr\left(YMY^T\right).
\end{equation} 
In order to avoid degenerate and non-unique solutions, the minimization is carried out subject to $\sum_{j=1}^p y_{\bullet j}=\vec{0}$ and $p^{-1}YY^T=I_{n'}$. Under these circumstances, for an eigenvalue decomposition $M=Q\Lambda Q^T$, the optimal embedding is given by
\begin{equation}
Y=\sqrt{p}\begin{pmatrix} q_{\bullet p-1}^T \\ \vdots \\ q_{\bullet p-n'}^T\end{pmatrix} \in \mathbb{R}^{n' \times p}.
\end{equation} 

\section{Visualization}
One of the biggest challenges in the areas of feature extraction and dimensionality reduction is to follow and understand the process from the input matrix $X$ to the (possibly dimensionality reduced) output matrix $Y$. Although the underlying mathematical frameworks are well motivated, the application resembles a black box system, in which the choice of parameters and/or the interpretation of the results often is the major difficulty. We present two new visualization methods that assist the interpretation after feature extraction or dimensionality reduction and give further insights into the structure of high dimensional data sets by visualizing important intermediate steps. Although both techniques are applicable in a universal context, they are especially useful in situations where the features or experiments allow a meaningful ordering. This is the case in many real-world applications, for instance when analyzing a genome or conducting a time series. 
\paragraph{Loading maps and their relatives}
While the mathematical elegance of a PCA is appealing, it has always been a major challenge to interpret the principal components. Because they are linear combinations of the original variables, it is a quite difficult task to assign a biological, physical or chemical meaning to them. To achieve more clarity, we propose to combine four theoretical measures that assist the interpretation with a visual component. Note that there are three fundamental terms in PCA: variables, observations and principal components. Usually, one is particularly interested in their mutual dependency \cite{Abdi2010}, especially in the importance of
\begin{itemize}
\item a variable for a principal component (loadings):
\begin{equation}\label{L0}
L_0\left(x_{i\bullet},y_{j\bullet}\right) = q_{ij}^2.
\end{equation}
\item a principal component for a variable (squared Pearson correlation):
\begin{equation}\label{L1}
L_1\left(y_{j\bullet},x_{i\bullet}\right) = r\left(y_{j\bullet},x_{i\bullet}\right)^2.
\end{equation}
\item an observation for a principal component: 
\begin{equation}\label{C0}
C_0\left(x_{\bullet j},y_{i\bullet}\right)=\frac{\left<q_{\bullet i},x_{\bullet j}\right>^2}{\lambda_i\left(p-1\right)}.
\end{equation}
$C_0$ describes the proportion of the variance of the $i$-th principal component that is caused by the $j$-th observation.
\item a principal component for an observation: 
\begin{equation}\label{C1}
C_1\left(y_{i\bullet},x_{\bullet j}\right)=\frac{\left<q_{\bullet i},x_{\bullet j}\right>^2}{\left\lVert x_{\bullet j}\right\rVert^2}.
\end{equation}
This is also known as the squared cosine and describes how much the \hbox{$i$-th} principal component contributes to the squared distance of the $j$-th observation.
\end{itemize}
Note that $\sum_{i=1}^n L_0\left(x_{i\bullet},y_{j\bullet}\right) = 1$ for all principal components $y_{j\bullet}$. This is the basis for a visual representation that we term a \emph{loading map}. For a particular principal component, a loading map displays the loadings of all variables as a stacked histogram, with the loadings sorted according to a given ordering. Because this type of representation is very space saving, it allows the parallel depiction of the loading maps of all principal components at once, thus giving within a single figure an immediate impression of the structure of the data set and the meaning of a particular principal component. Variables or groups of variables which are of special interest to the analysis (e.g., clusters of variables that are known beforehand) may then be colored to show their positions in the loading maps. After appropriate scaling, $L_0$ can be generalized to produce ICA-loading maps in a straightforward manner.

\begin{figure}[tbp]
  \centering
%  \subfigure[Loading matrix.]{
%  \raisebox{1.8cm}{
\parbox{0.45\textwidth}{\Large
  $$\left(\begin{array}{cccc}0.25 & ~0~ & 0.1 & ~0.7 \\0.25 & ~0~ & 0.5 & ~0.1 \\0.25 & ~1~ & 0.3 & ~0.1 \\0.25 & ~0~ & 0.1 & ~0.1\\
  \end{array}\right)$$
}\hfil  
\parbox{0.45\textwidth}{
%  }
  %\includegraphics[width=2.6cm]{figure1a.pdf}
%  \label{LMEX1}
%  }
%  \hspace{1cm}
%  \subfigure[Loading map.]{
  \includegraphics[width=5.6cm]{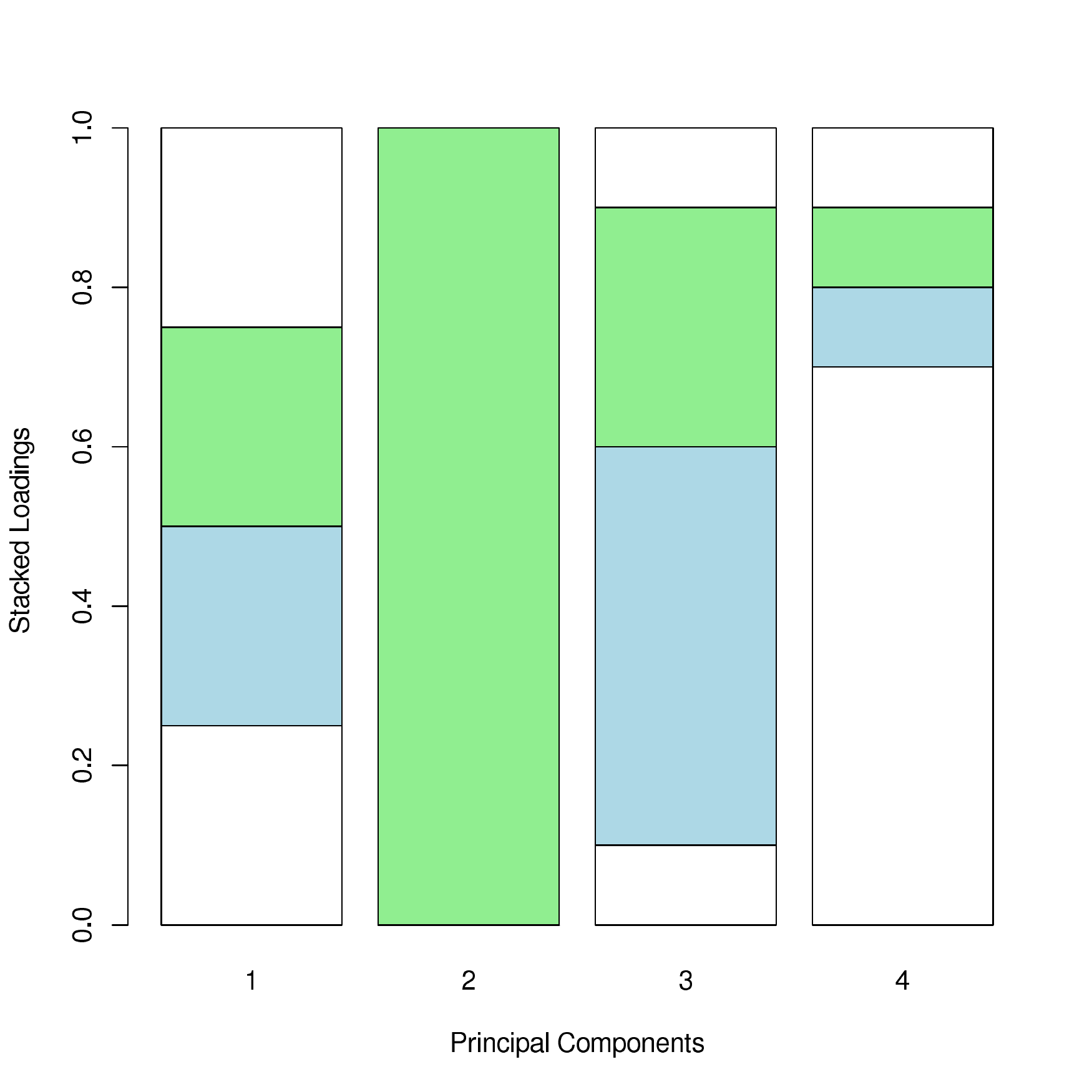} 
  }
  
\parbox{0.45\textwidth}{\centering
(a)  Loading matrix}\hfil  
\parbox{0.45\textwidth}{\centering
  (b) Loading map
  }
%  \label{LMEX2}
%  }
  \caption{Conversion of a loading matrix to a loading map. Figure~\ref{LMEX}(a) shows a hypothetical $4\times 4$ loading matrix and Figure~\ref{LMEX}(b) the corresponding loading map. In this example, the variables $X_2$ and $X_3$ have been colored in blue and green, respectively, in order to show their positions in the map.}
  \label{LMEX}
\end{figure}

Consider the following illustrative example: We carry out a PCA on a $4\times p$ matrix, $p>4$, and collect all loadings in a loading matrix $L_0=\left(L_0\left(x_{i\bullet},y_{j\bullet}\right)\right)_{1\leq i,j \leq 4}$ that is subsequently converted to a loading map. Figure \ref{LMEX} demonstrates this process for a hypothetical loading matrix. Further down, we will give a detailed example where the variables are given by genes, the ordering is given by their sequence in the genome and the colors represent different functional clusters. 

In fact, summation over the first component in any of these measures yields $1$, so the same approach works for the other three measures (Eqs. \ref{L1}-\ref{C1}) as well. This means we obtain a class of four visualizations that are very similar in appearance, yet they show completely different aspects of the analysis and complement each other so as to assist the interpretation of principal components.

\paragraph{$l$-neighborhood evolution matrices}
Our second contribution to the elucidation of important steps in the process of dimensionality reduction takes the form of what we termed \emph{$l$-neighborhood evolution matrices ($l$-NEMs)}. $l$-NEMs are a helpful extension to the use of classical dot plots, which have traditionally been used to compare two sequences, especially biological sequences like DNA or protein sequences. For $l$ colors $c_1,\ldots,c_l$, we extend this idea to the comparison of points in $n$-dimensional Euclidean space by coloring a cell $\left(i,j\right)$ of a $p\times p$ grid with color $c_k$, if $x_{\bullet j} \in \mathbb{R}^n$ is the $k$-nearest neighbor of $x_{\bullet i} \in \mathbb{R}^n$. The similarity between two points can be determined by an arbitrary distance function, e.g., the Minkowski distance, correlation distance or any other distance function. If the distance function is symmetric and we consider all points as neighbors whose distance is smaller than a fixed radius $R$, then the neighborhood evolution matrix is symmetric as well (however not necessarily the colors of the dots). Note that this need not be the case for a fixed number of neighbors. The actual number of neighbors displayed is variable and allows for an adaptive view that shows how the neighborhood develops (hence the name $l$-NEM).

\begin{figure}[tbp]
  \centering
  \subfigure[3-NEM.]{
  \includegraphics[width=5cm]{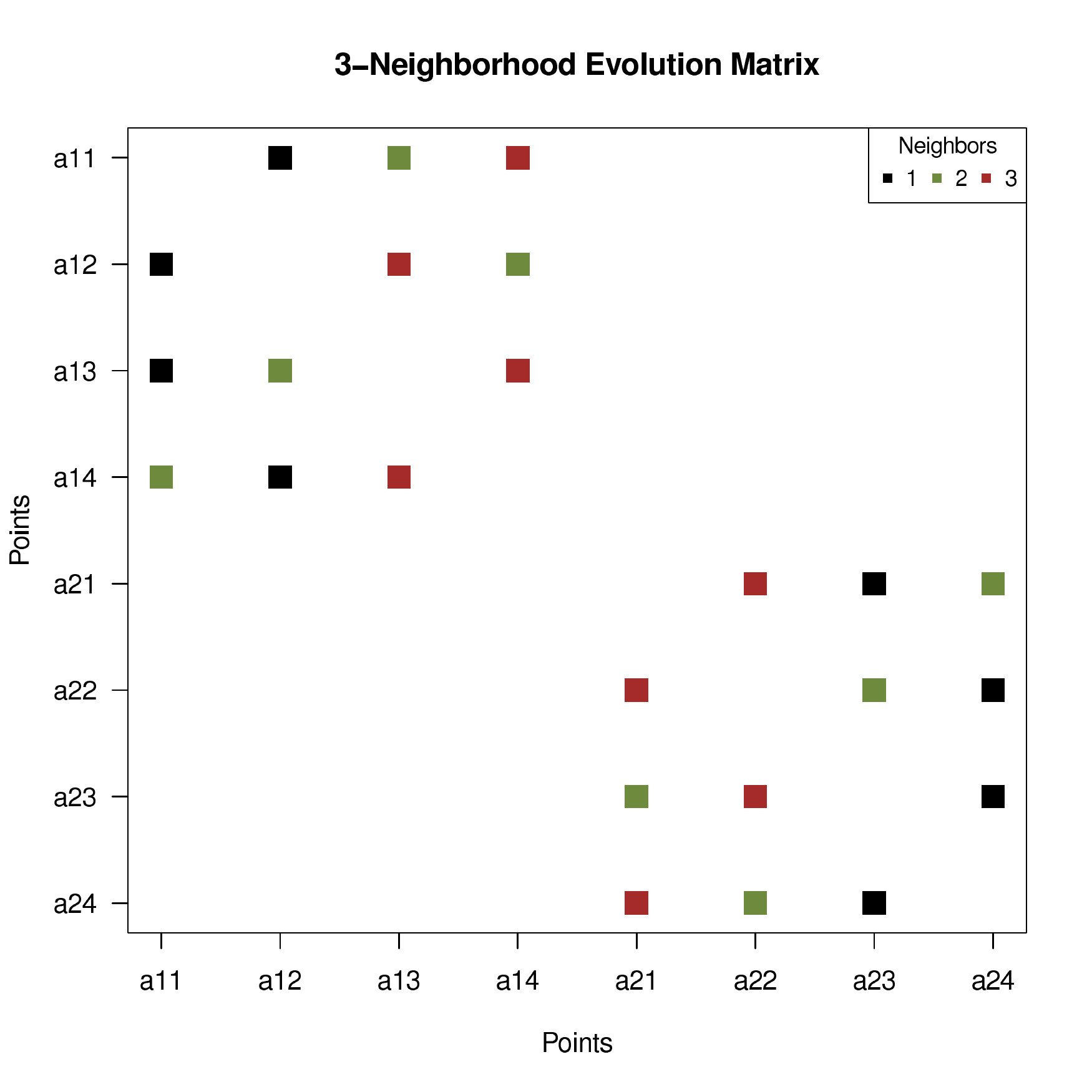}
  \label{3NEMEX}
  }
  \hspace{1cm}
  \subfigure[5-NEM.]{
  \includegraphics[width=5cm]{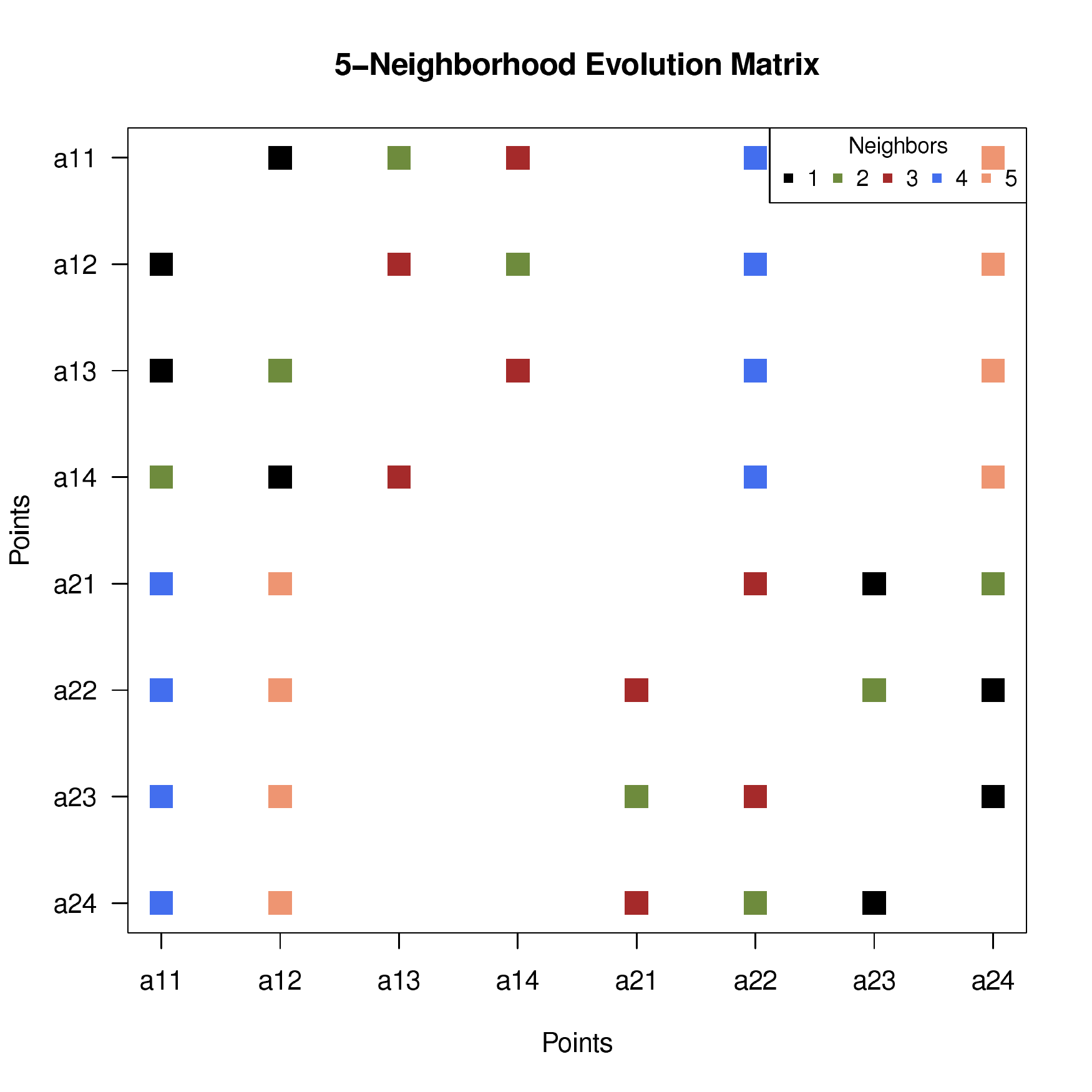} 
  \label{5NEMEX}
  }
  \caption{$l$-neighborhood evolution matrices ($l$-NEMs). The figure shows a comparison between a $3$-NEM (a) and a $5$-NEM (b) originating from two clusters in $n$-dimensional space. While the $3$-NEM shows a restriction of the neighborhood to the individual clusters, the $5$-NEM indicates a `cluster overflow'.}
  \label{NEMEX}
\end{figure}

The coloring scheme is taken from ColorBrewer \cite{ColorBrewer}, which offers color sets to allow easy differentiation of the neighbors.
%these qualitative data.

As an example, consider two compact clusters $A_1=\left(a_{11},\ldots,a_{14}\right)$ and $A_2=\left(a_{21},\ldots,a_{24}\right)$ in $n$-dimensional space (e.g., time points) whose single linkage distance is larger than any distance between two points within the clusters. In this case, we obtain a $3$-NEM that is similar to that in Figure~\ref{3NEMEX}. The extension to a $5$-NEM leads to a `cluster overflow' and the neighborhood extends to the other cluster (see Figure~\ref{5NEMEX}). 

$l$-NEMs are of special use for confirming the number of neighbors that one intends to use in LLE. Originally, Saul and Roweis proposed increasing the neighborhood size until the neighborhood graph is strongly connected \cite{Saul2003}. However, in situations like the one mentioned above, we may have strong connectivity only on account of there being a very small bottleneck. This is easily discovered using $l$-NEMs.

\section{SpRay}
All of the feature extraction/dimensionality reduction methods described so far, as well as the eigenvalue decomposition (EVD) and the singular value decomposition (SVD), have been implemented within our interactive visualization framework SpRay \cite{Dietzsch2009}. SpRay is a comprehensive, flexible and extendable software system that combines refined visual exploration with statistical methods to provide a unique visual analytics environment that proved to be particularly successful for the visual exploration of gene expression data. Different visual components like the parallel coordinates plot (PCP), the table lens and the scatterplot matrices are easily accessible in a linked view setting. SpRay uses a module-based approach and connects different processing modules to a pipeline that realizes the data flow. Along the pipeline, each module adds specific information to the data model and the visual components at the very end of the pipeline are able to access all of the information produced by the modules as and when required.

\begin{figure}[tb]
  \centering
  \includegraphics[width=10cm]{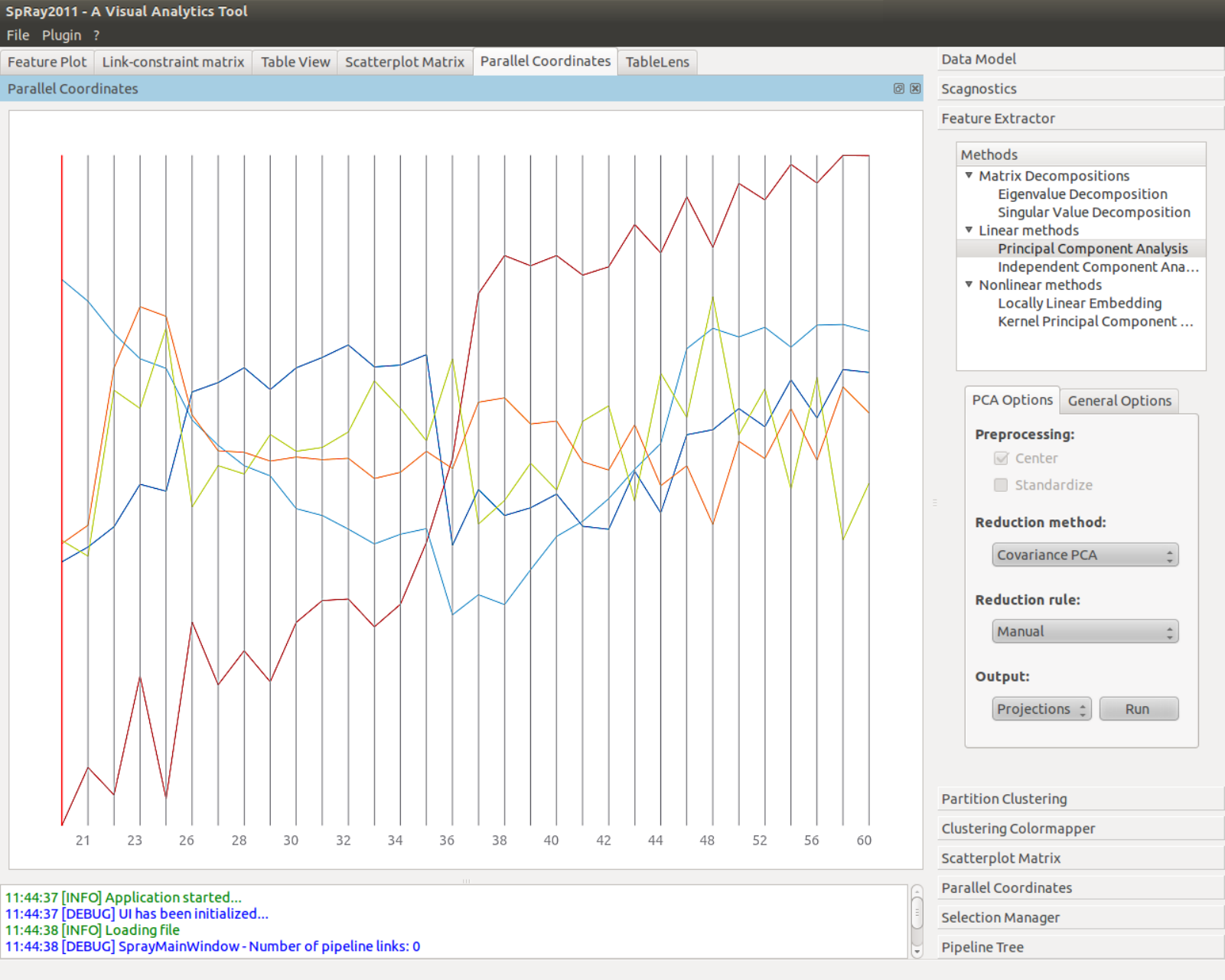} 
  \caption{The graphical user interface of SpRay. The options are accessible through a toolbox on the right, the visualizations through a tabbed navigation at the top of the main window.}
  \label{SPRAYGUI}
\end{figure}

PCA, ICA and LLE are integrated into a single module named `Advanced Statistical Methods Library' (ASML) that allows for fast switching between the different methods, thus permitting an immediate comparison of the results. The desired output dimension and the underlying mode of operation (i.e., reduction of rows (features) or columns (experiments)) can be chosen in an interactive manner. Besides the basic procedures, the implementation includes several variants, extensions and modifications of the original algorithms. These include:

\begin{enumerate}
 \item Principal Component Analysis: All four measures that assist the interpretation of the principal components (Eqs. \ref{L0}-\ref{C1}); five different methods that recommend how many principal components should be kept; and a modification that uses correlation matrices instead of covariance matrices. 
 \item Independent Component Analysis: A cumulant-based and a negentropy-based approach to maximize non-Gaussianity; high and low pass filters to preprocess the data; and three contrast functions to approximate negentropy \cite{Hyvaerinen1999}.
 \item Locally Linear Embedding: Computation of strongly connected components in the neighborhood graph \cite{Tarjan1972}; restriction to a specific submanifold in the disconnected case; and a version of the algorithm assuming $X$ to be a distance matrix \cite{Saul2003}.
\end{enumerate}

Figure~\ref{SPRAYGUI} shows the graphical user interface of SpRay while analyzing a test data set. As can be seen from the selected options in the menu to the right, the parallel coordinates plot, in this case, shows the results after applying a reduction to five output dimensions with subsequent principal component analysis.  

After processing the data, the loading maps as well as neighborhood evolution matrices can directly be generated on demand by the user within SpRay. For this, \verb R  is used, which is directly embedded in SpRay.
In addition, SpRay writes the results obtained to a file in an \verb R -compatible format. This technique allows for a seamless exchange between the two systems so that loading maps as well as neighborhood evolution matrices can be easily produced in \verb R  with adapted parameters.

SpRay is available for the community at \url{http://www-ps.informatik.uni-tuebingen.de/sprayWeb/}.

\section{Application to expression data}
To exemplify our visual analytics approach for dimension reduction we applied it to two different large-scale expression data sets. In the first example, reduction is applied to the genes, while in the second example, we demonstrate how the loading maps can be applied to a principal component analysis of experiments.

\subsection{Time series expression data}
For the first example,
we use a data set from a large-scale transcriptomic study of the soil bacterium {\em Streptomyces coelicolor} grown under phosphate limited conditions \cite{Nieselt2010}. From 20 hours after inoculation, samples were collected every hour up to 44 hours, and every two hours afterwards, resulting in a total of 32 time point samples. Global gene expression profiles were acquired  for each sample using custom designed Affymetrix GeneChips \cite{Battke2011}. The study revealed a very dynamic expression landscape as the organism undergoes a major metabolic switch from exponential growth to antibiotic production. 
 We applied our dimension reduction methods to the matrix with 7893 variables (the genes) and 32 columns (the time points).

\begin{figure}[tb]
\centering
\includegraphics[width=0.49\textwidth]{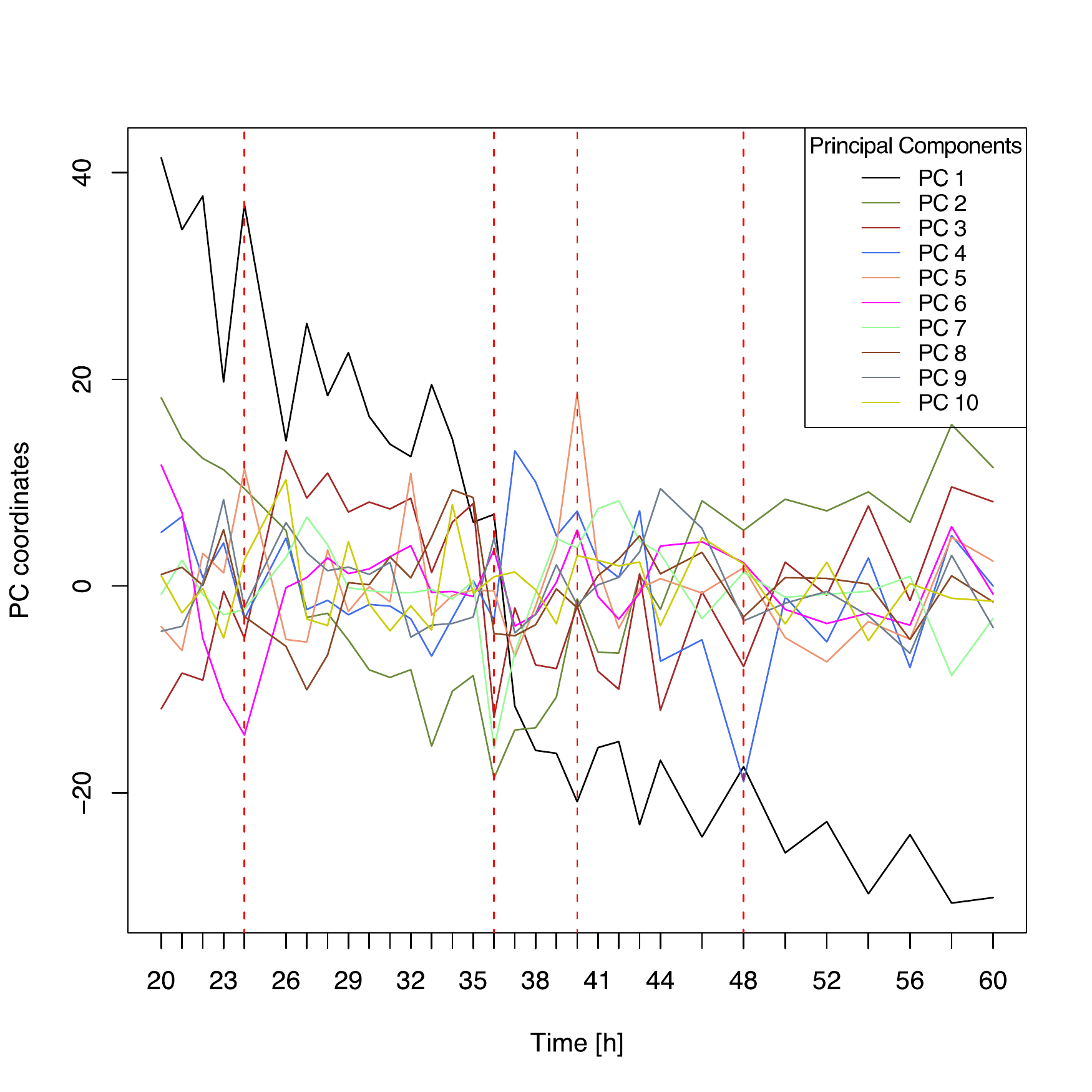}
%\hspace*{0.05cm}
\includegraphics[width=0.5\textwidth]{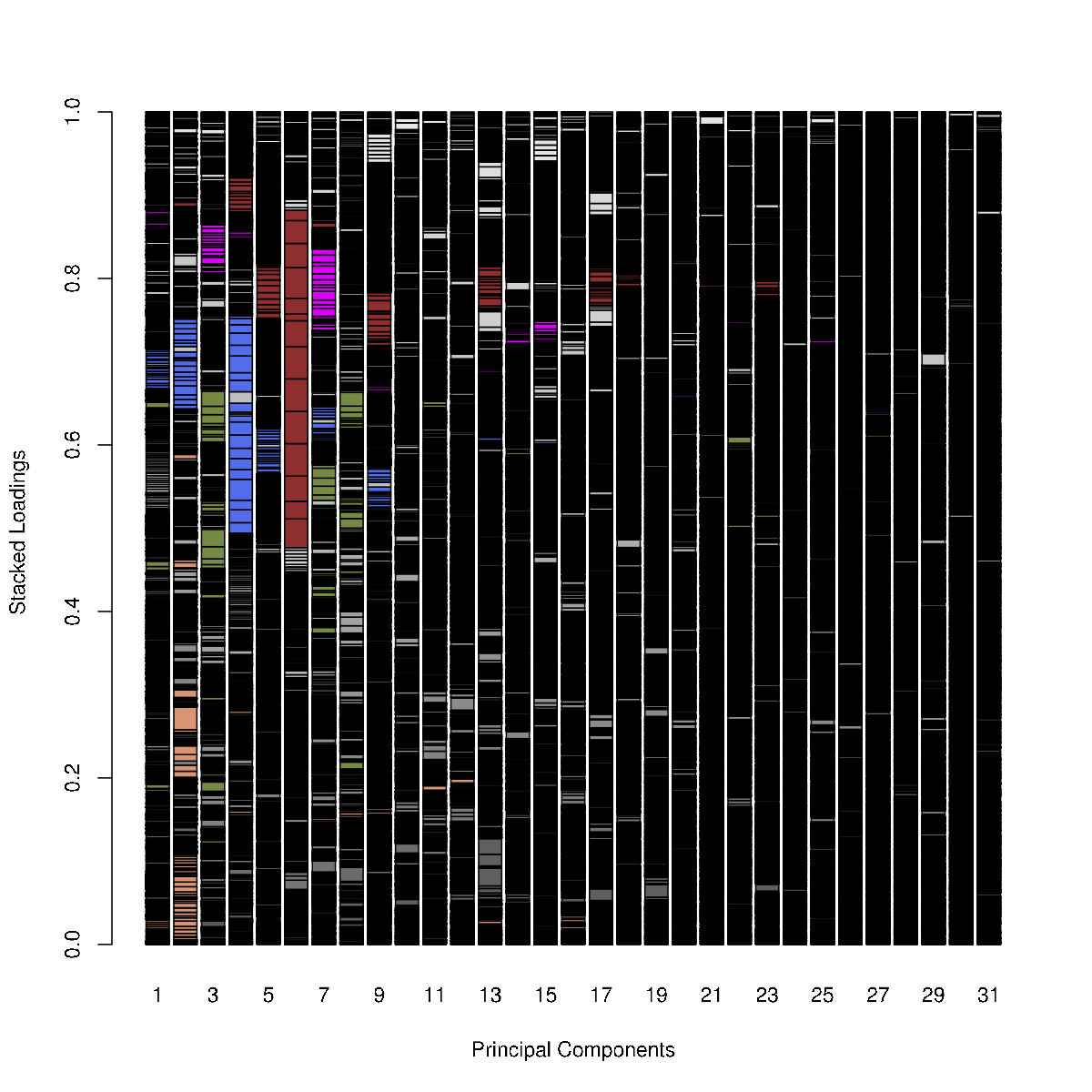}
\caption{Principal component analysis (PCA) and genomic loading map of a time series microarray experiment in the bacterium {\em S.~coelicolor}. Here as a result, the profile plot of the first 10 principal components (left) is shown. The red dashed lines indicate four different important biological time points of expression changes.
The genomic loading map on the right visualizes the PCA results in their genomic and functional context. For each component the stacked histogram of the loadings of all genes, that are sorted by their genomic location is shown. Selected gene groups with common functional annotation are color labelled. Red: genes of the Cpk cluster, blue: genes of the Act cluster, pink: genes of the Red cluster, green: phosphate dependent genes, orange: developmental genes.
}\label{fig:PCA}
\end{figure}

We first computed the eigenvalue spectrum which revealed that the eigenvalues of this time series matrix are rather homogeneous. This result attests to the large structural complexity of this data which has a variety of underlying signals. 
Figure~\ref{fig:PCA} shows the parallel coordinate plot of the first ten principal components, capturing 90\% of the data's variance. 

The profiles of the principal components reflect the dynamics of gene expression of this data with the most drastic change at 36 hours. In addition, at least three further time points, at 24, 40 and 48 hours, can be identified where at least one of the principal components shows a large increase or decrease in its coordinates. The authors in \cite{Nieselt2010} have identified these time points to signal important onsets of regulation during the growth of the bacterium.

However, in order to gain greater insight in particular into the classes of genes that are responsible for the signals, we computed the four measures as introduced in Eqs.~\ref{L0}-\ref{C1}. We started with the analysis of the loadings ($L_0$, Eq.~\ref{L0}) 
and their visualization by {\em genomic} loading maps (Figure~\ref{fig:PCA}).
In a genomic loading map each principal component is visualized in terms of a stacked histogram of the loadings of each gene, with the genes ordered according to their chromosomal position. Since the loadings sum up to 1 for each PC, this visualization gives an immediate overview of which genomic locations are relevant for the PC.
In addition, we colored genes with a common functional annotation. The visualization in Figure~\ref{fig:PCA} clearly shows that no specific gene group dominates the first PC. PC~1 basically represents the signal of the global trend and it does not reveal local individual signals of specific functional relevance. The loading map also shows large loadings appearing in chromosomal clusters, which clearly illustrates the common regulation of genes in the same genomic neighborhood, as is commonly observed in bacteria. 

\renewcommand{\subfigcapmargin}{10pt}
\begin{figure}[tbp]
  \centering
  \subfigure[Loadings ($L_0$ values) of genes along chromosome. Genes with loadings above the three-fold standard deviation are colored. One-fold, two-fold and three-fold standard deviations are denoted by grey horizontal lines.]{
    \label{SYSMO_AUSWERTUNG_F199_HK6_Loadings}
    \includegraphics[width=0.45\textwidth]{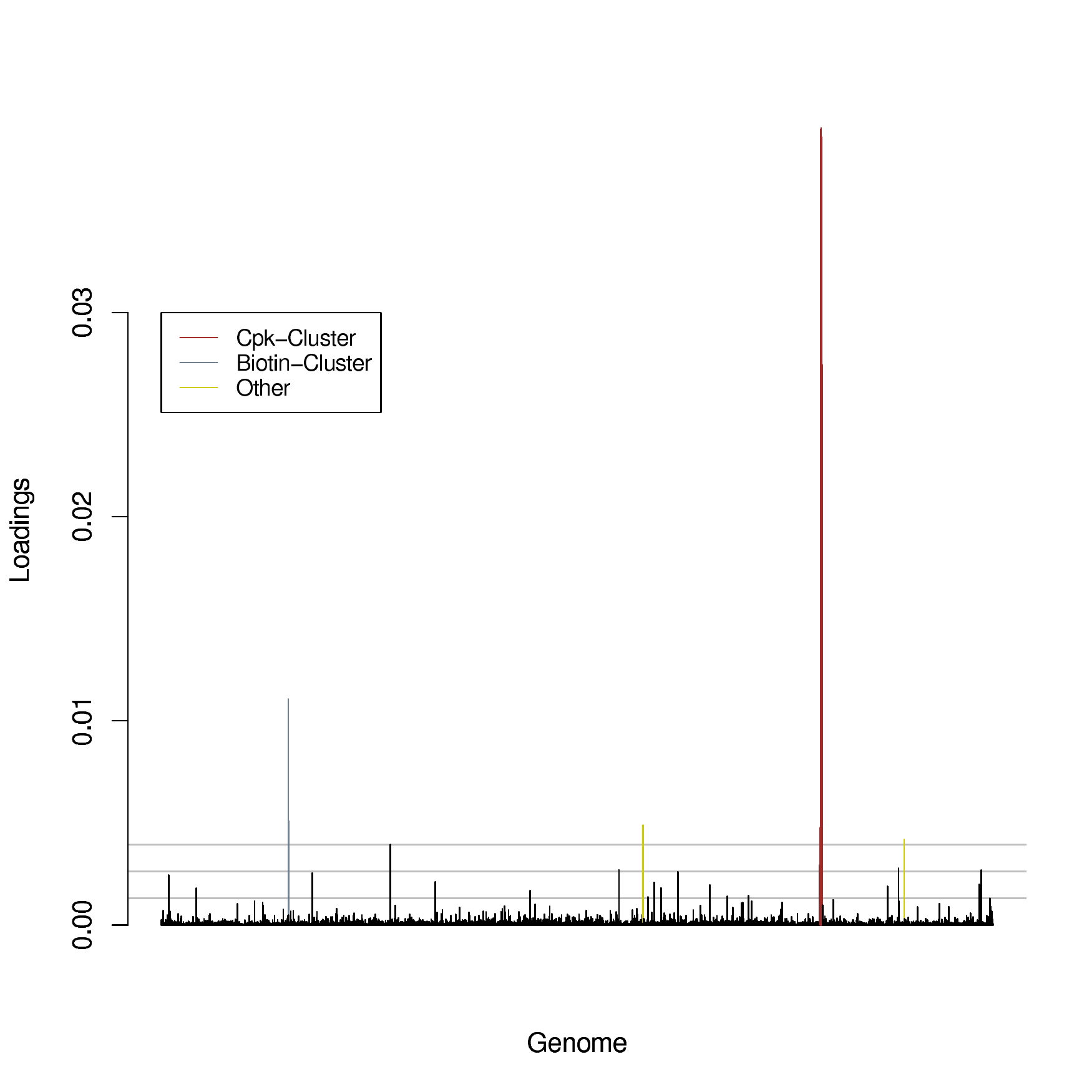} 
  }
  \subfigure[Expression profiles of genes with $L_0$ values above the three-fold standard deviation. The black profile is the mean profile of this group.]{
    \label{SYSMO_AUSWERTUNG_F199_HK6_Cluster}
    \includegraphics[width=0.45\textwidth]{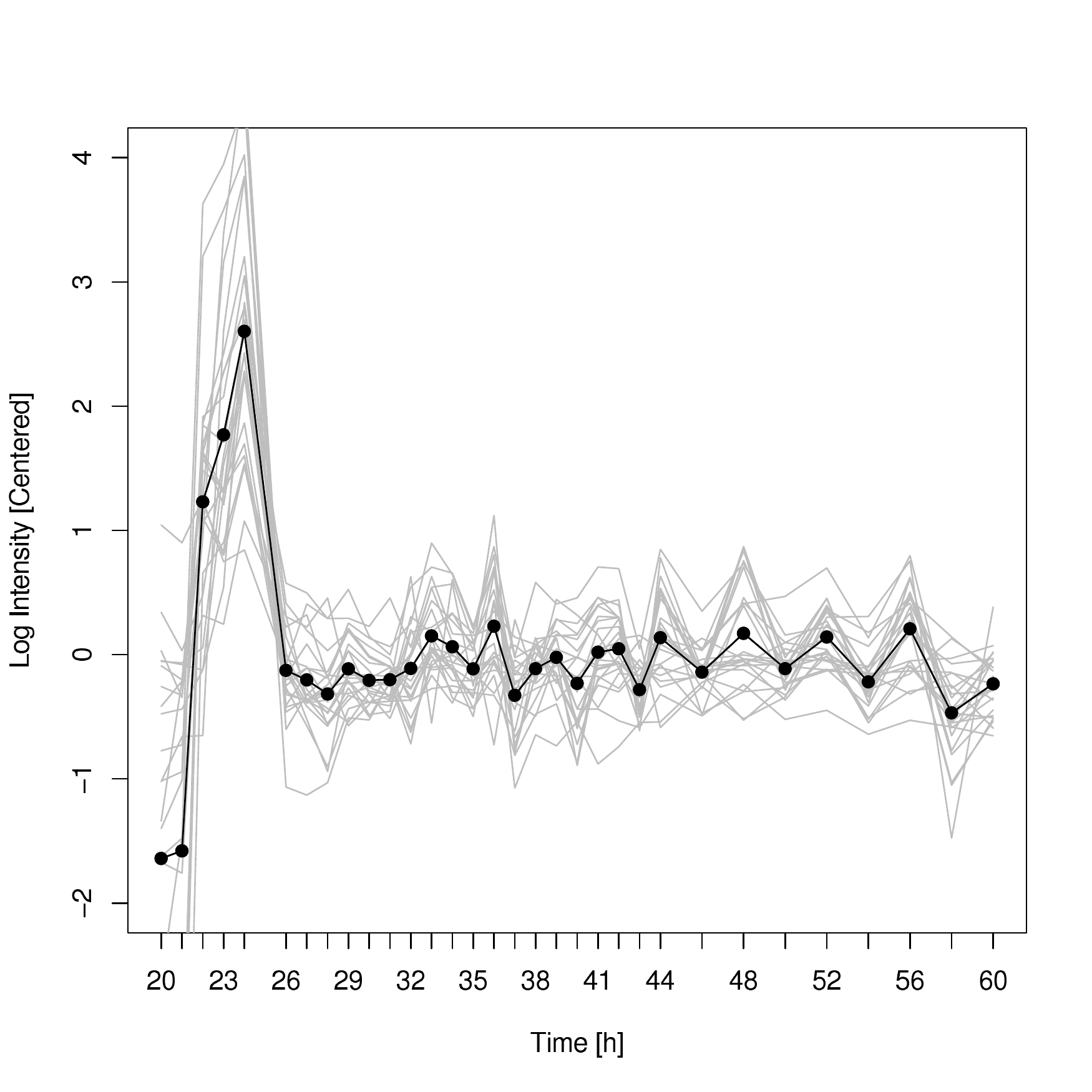} 
  }  
  \subfigure[Time dependence profile ($C_0$ values) of PC~6.]{
    \label{SYSMO_AUSWERTUNG_F199_HK6_TD}
    \includegraphics[width=0.45\textwidth]{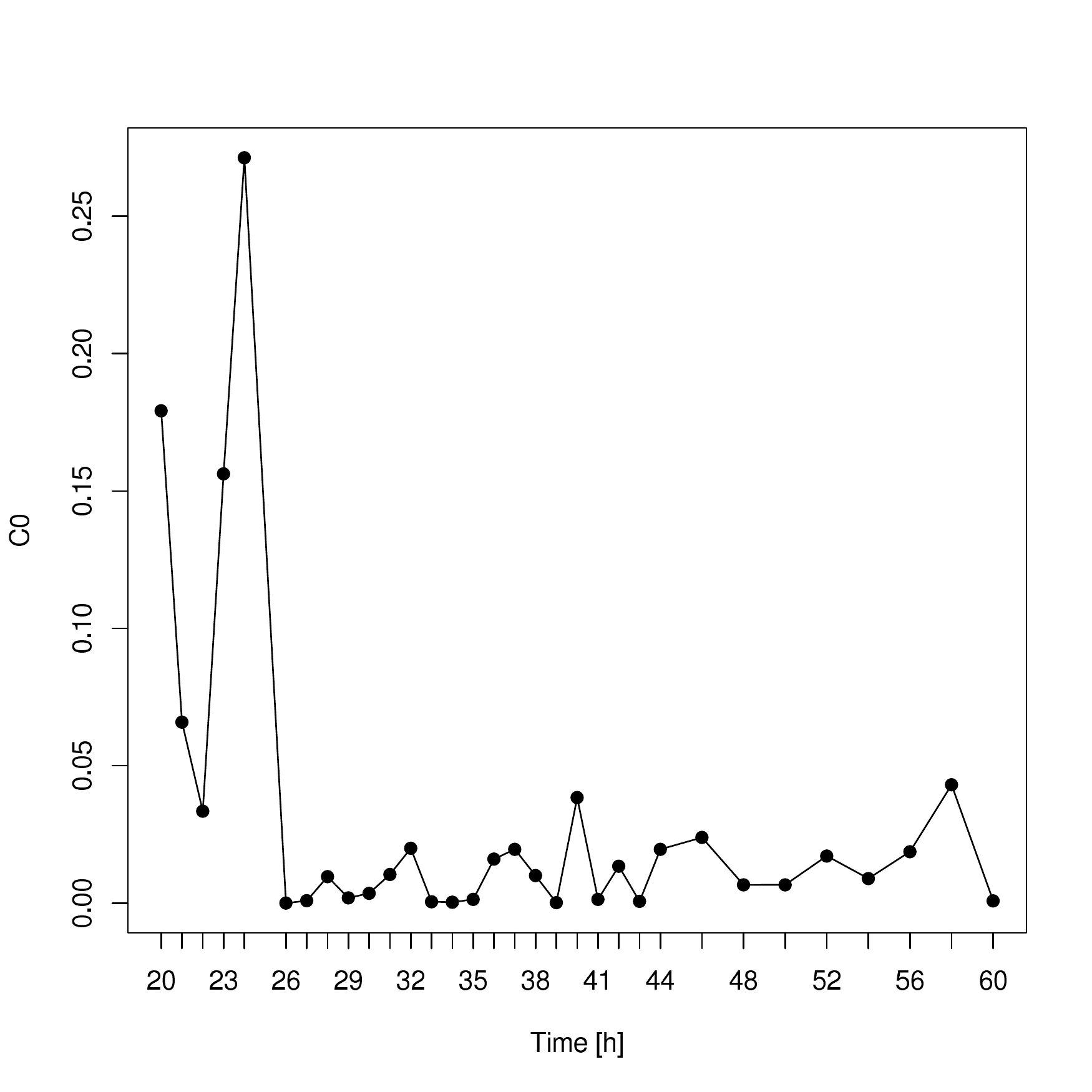} 
  }
  \subfigure[$L_1$ values of genes of Cpk cluster (in grey). In black the mean $L_1$ values of this cluster are plotted.]{
    \label{SYSMO_AUSWERTUNG_F199_HK6_CPK}
    \includegraphics[width=0.45\textwidth]{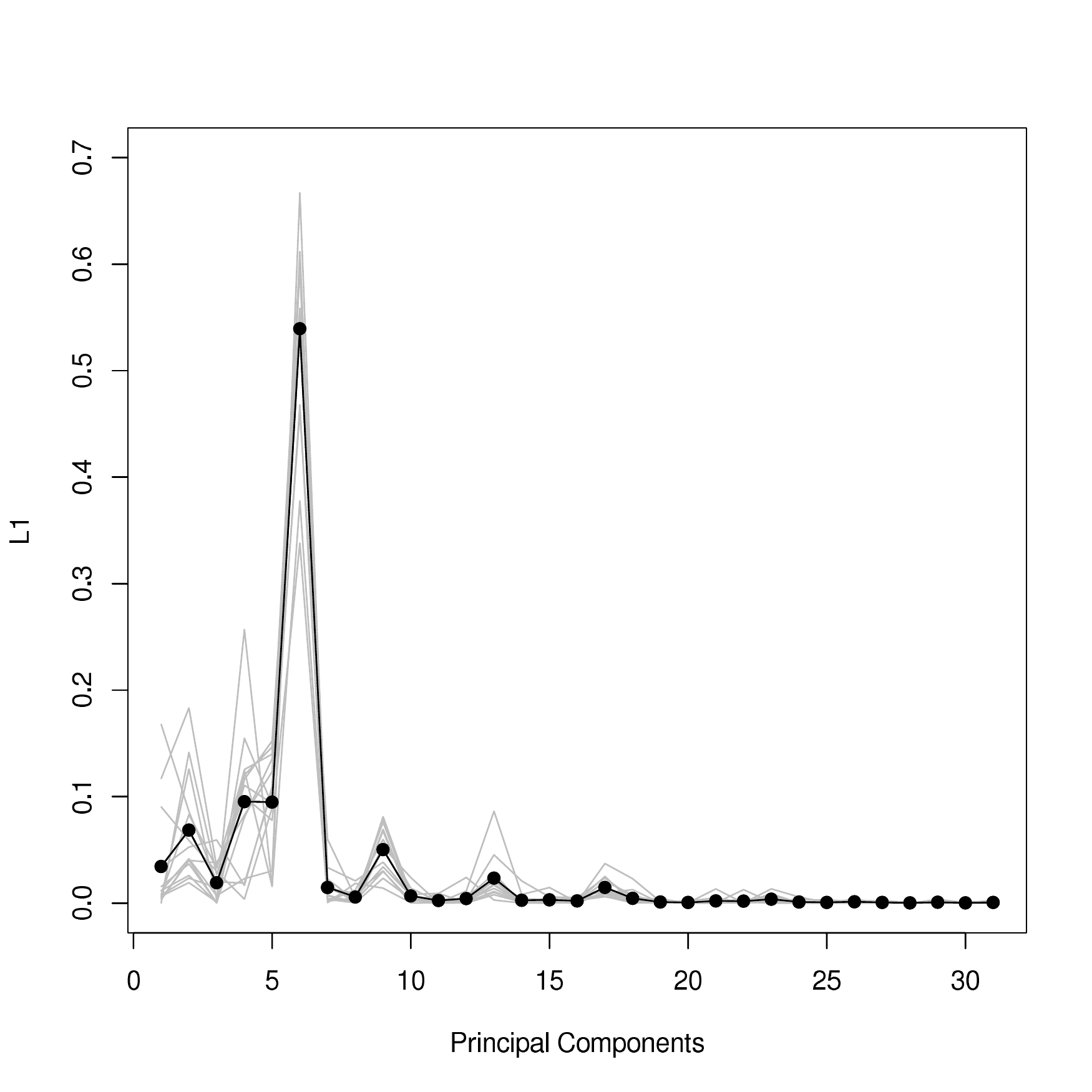} 
  }
  \caption{Complete characterization of the principal component 6 using the measures $L_0$, $C_0$ and $L_1$.}
  \label{fig:HK6}
\end{figure}

Loading maps offer a complete and at the same time compact view of all components, and therefore allow a direct comparison of all components. Genes with large loadings are clearly identifiable, while those with very small loadings are almost invisible.

In order to get an even clearer picture of individual components with high structural value, i.e., those that have few large loadings and most loadings close to zero, we took a closer look at the sixth principal component.
The visualization of its detailed characterization using the $L_0$, $C_0$ and $L_1$ measures is depicted in Figure~\ref{fig:HK6}.
From the visualization of the loadings of all genes in this PC, we find four clearly separated regions of genes, among them a group with very large loadings, whose corresponding profiles are very homogeneous, nicely showing the early up-regulation of these genes before 26h (Figure~\ref{fig:HK6}b).
Furthermore, we see that PC~6 encapsulates this early expression event (Figure~\ref{fig:HK6}c) almost exclusively
and that the cpk genes with very large loadings are also mostly captured by this PC (Figure~\ref{fig:HK6}d).

\begin{figure}[tb]
\centering
\includegraphics[width=0.495\textwidth]{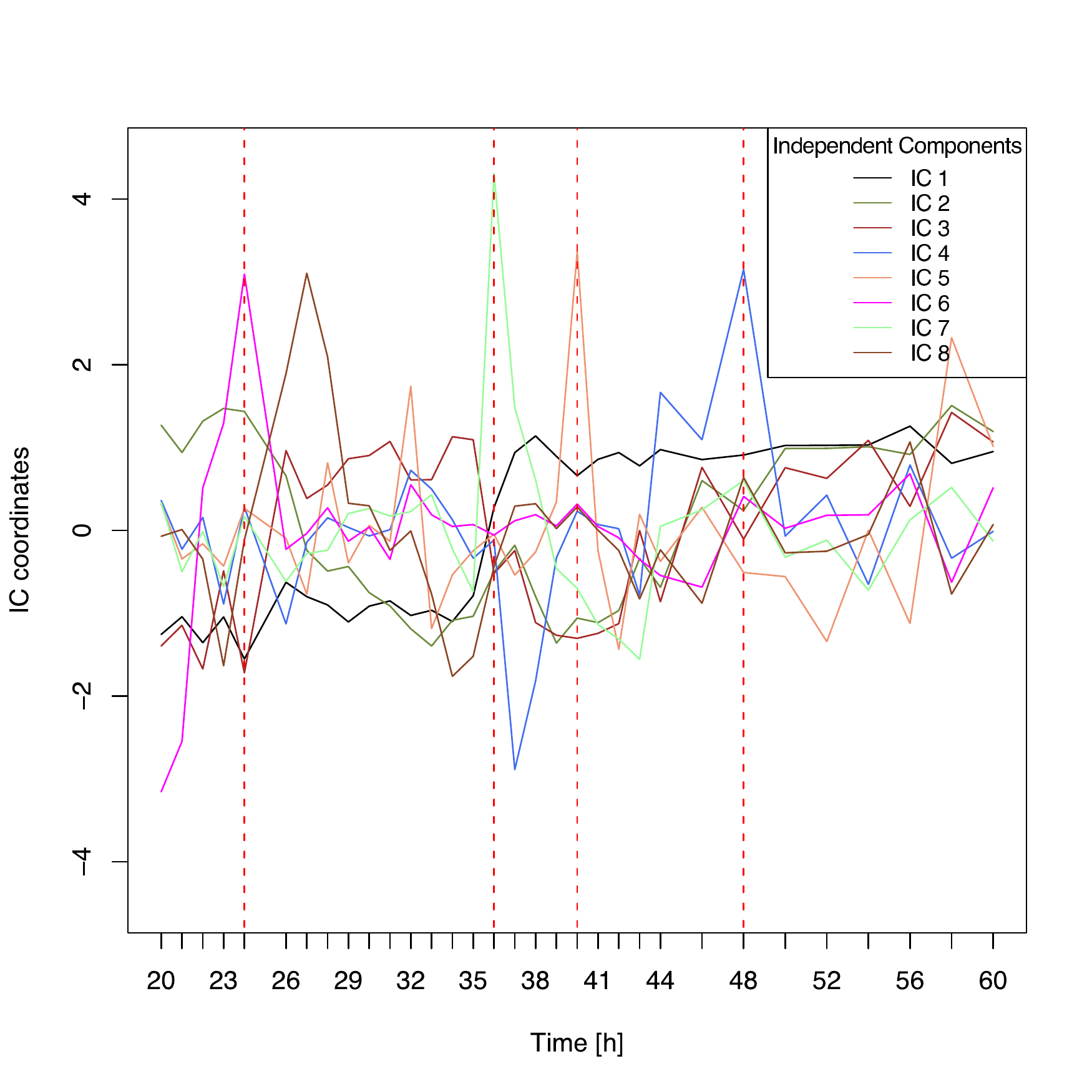}
\includegraphics[width=0.495\textwidth]{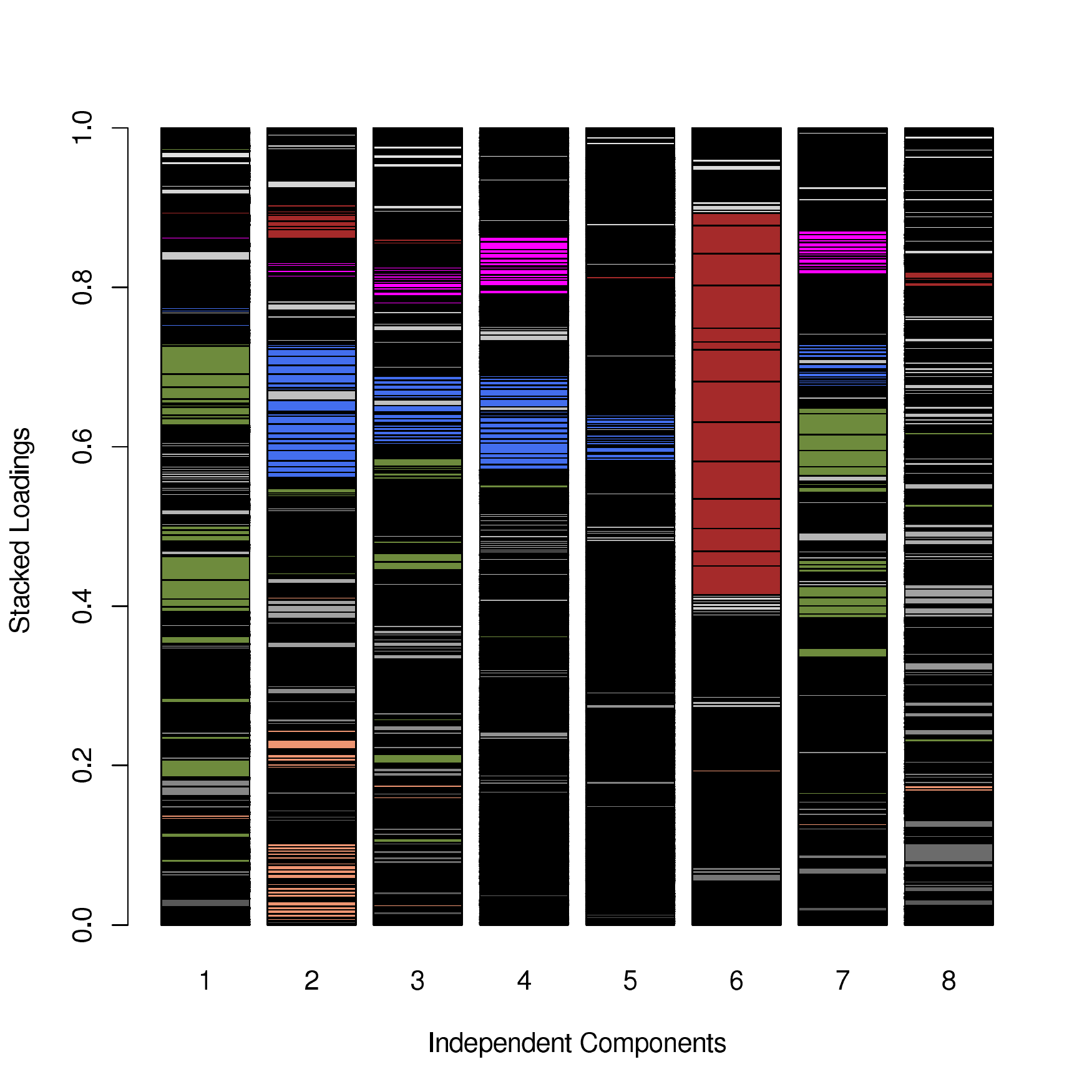}
\caption{Profile plot of independent components as a result of the independent  component analysis (left) and genomic loading map (right). The same color map as in Figure~\ref{fig:PCA} was applied.
}\label{fig:ICA}
\end{figure}

The application of independent component analysis to these data revealed the importance and complementarity of this method to PCA for this type of data. The profile plot of the first eight independent components (Figure~\ref{fig:ICA} left) clearly shows the ability of the ICA to detect important signals by individual independent components (IC~6 for time point 24h, IC~7 for 36h, IC~5 for 40h, and IC~4 for 48h). In contrast to the first PC, the first IC represents the change from primary to secondary metabolism at 36h, marked by an almost homogeneous coloring of all those genes that have large loadings, as can be seen in the equivalent IC genomic loading map (Figure~\ref{fig:ICA} right). Interestingly, the expression profiles of these genes are very similar to the profile of IC~1.

\begin{figure}[tb]
\centering
\includegraphics[width=0.495\textwidth]{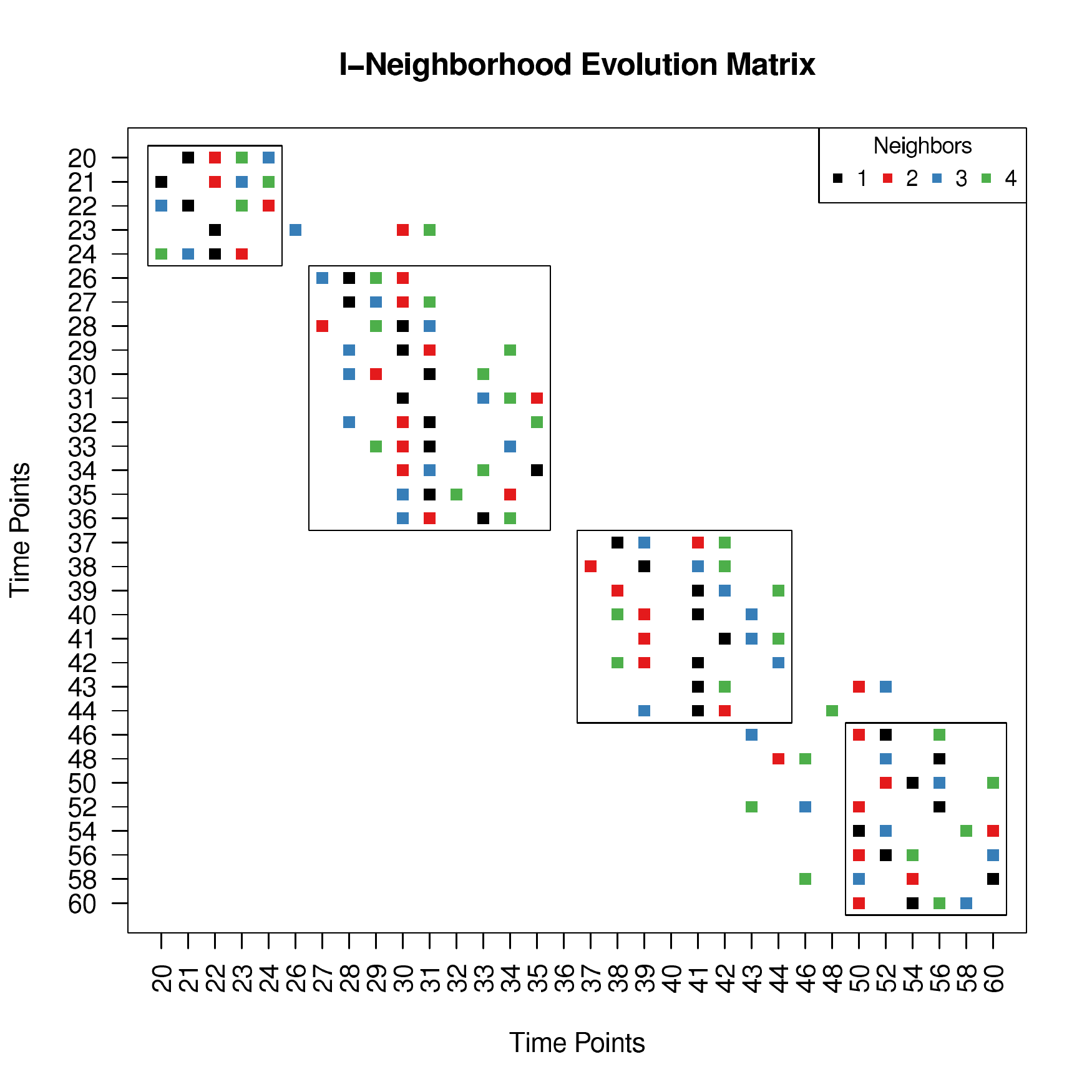}
\includegraphics[width=0.495\textwidth]{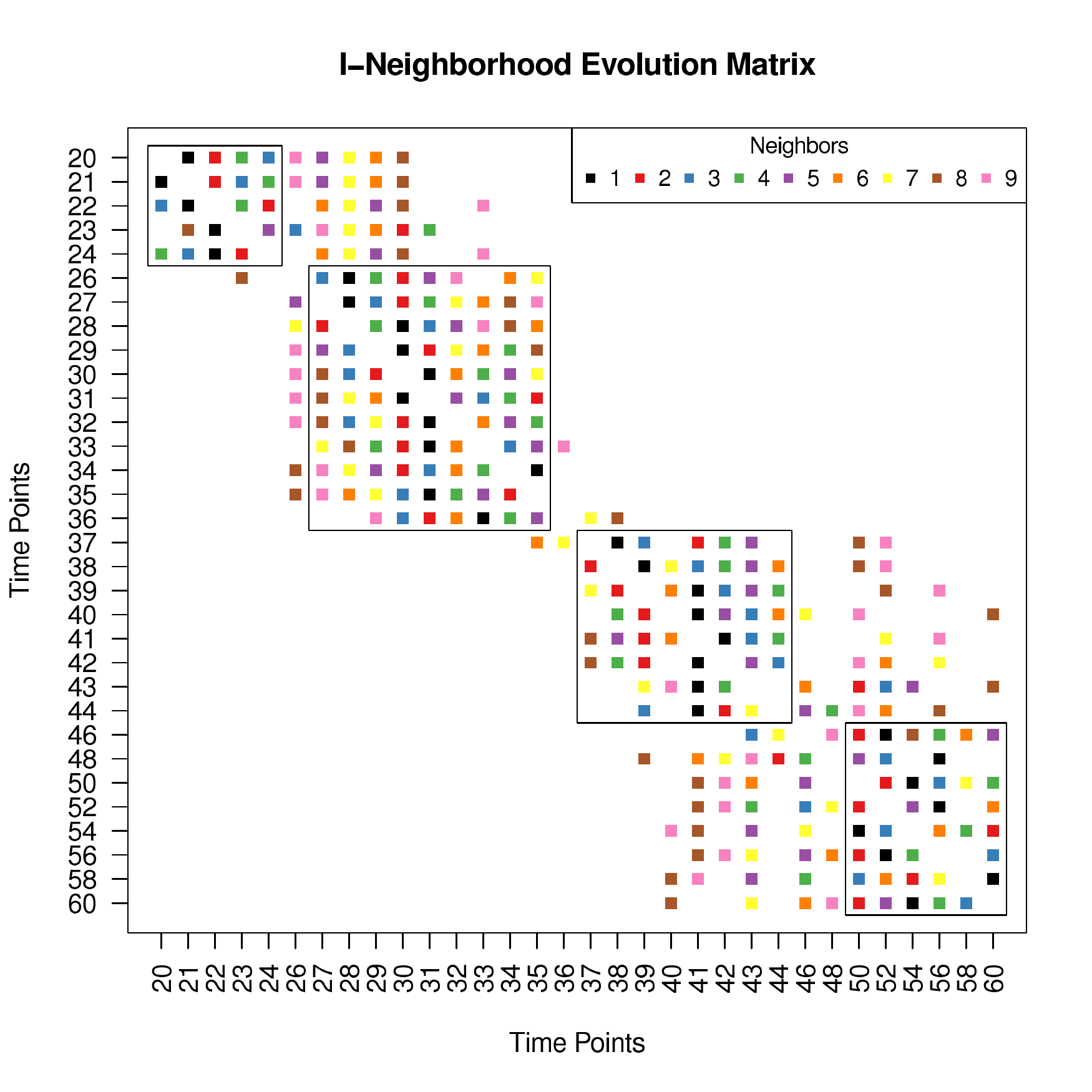}
\caption{Visualization of $l$-neighborhood evolution matrices of the expression time series in {\em S.~coelicolor}. On the left $l=4$, on the right $l=9$ neighbors were chosen, using the Euclidean distance function. The colors were chosen with ColorBrewer \cite{ColorBrewer}.
}\label{fig:LLE-F199}
\end{figure}

Finally, we applied LLE to this time series. To our knowledge the LLE method has so far not been applied to results from a time series expression experiment. In order to get a first impression of the neighborhood, we started by producing neighborhood evolution matrices for $l=4$ and $l=9$ neighbors (Figure~\ref{fig:LLE-F199}). They clearly support our time pattern hypothesis, that already few neighbors suffice to detect a clear block structure, and where the respective time points are nicely related to important biological time points during the growth of the bacterium. In comparison to the $4$-NEM, it is easy to see that the choice of 9 neighbors leads to a far better connected neighborhood graph because the four blocks (time clusters) start to interconnect. In Figure~\ref{fig:LLE-F199}, this is indicated by dashed rectangles that consist mainly of distant neighbors. Note also how distant neighbors tend to be offside the diagonal.

\subsection{Trisomy Expression Profiling}
In the second example, we used data from a genome-wide expression screen of cells with a normal karyotype and with trisomies of human chromosomes 13 and 21 \cite{Altug2008}. Using Affymetrix microarray technology, gene expression in amniocytes (AC) from pregnancies with a normal karyotype and with trisomies of human chromosomes 13 and 21 was measured. For each of the three cell types, three replicates were used. From the normalized expression data we extracted all genes with a minimum variance of 0.7 across the experiments. We then conducted a principal component analysis on the matrix with 9 variables (the samples) and 1911 observations (the genes).

\begin{figure}[tb]
\centering
\includegraphics[width=0.6\textwidth]{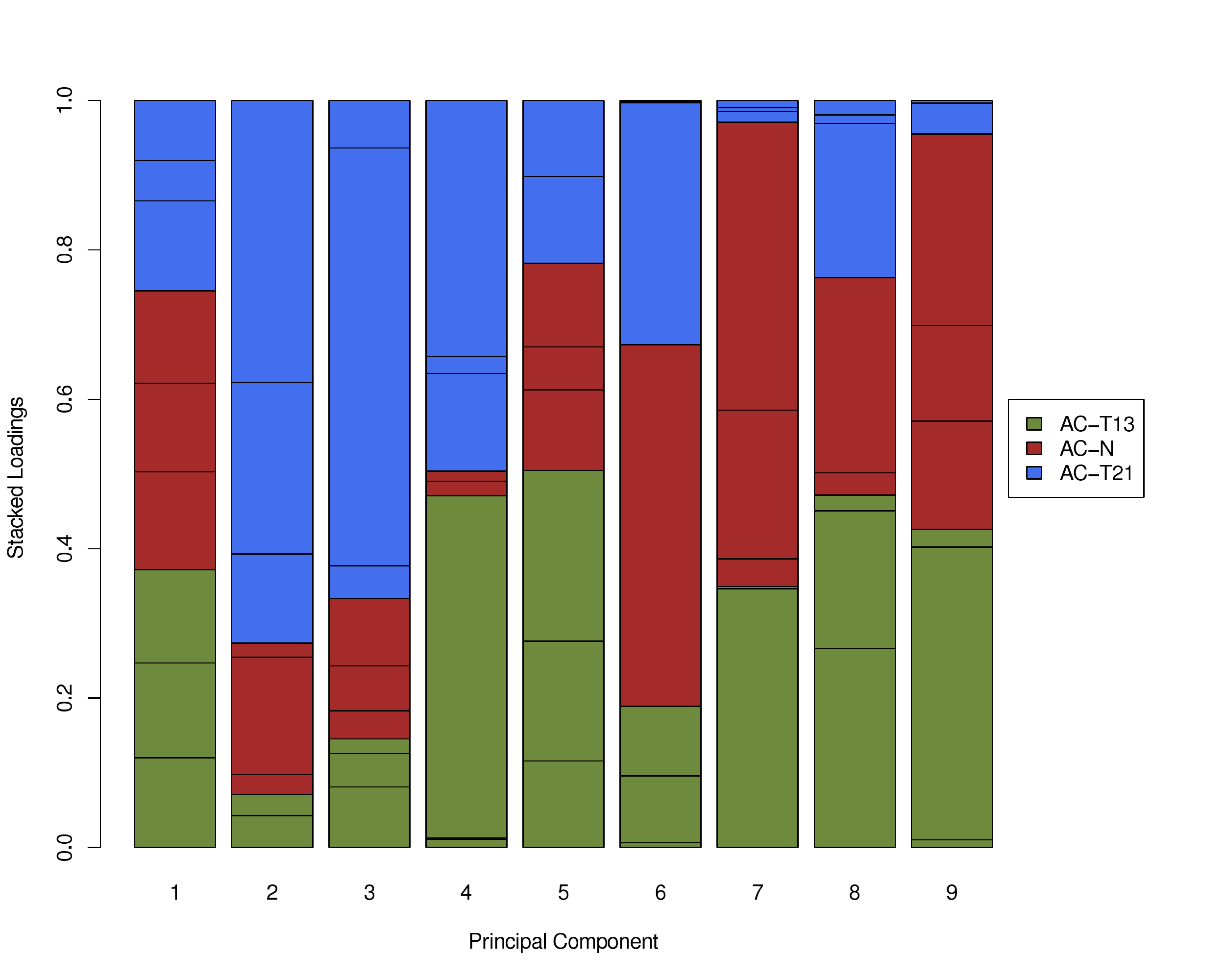}
\caption{Sample loading map of a microarray experiment comparing expression of human cells with autosomal trisomies with normal karyotype. 
The loading map visualizes the PCA results in their cell type context. For each component the stacked histogram of the loadings of all samples, that are sorted and grouped by their cell type is shown. AC-N: amniocyte cells with normal karyotype, AC-T13: amniocyte cells with trisomy 13, AC-T21: amniocyte cells with trisomy 21.
}\label{fig:PCASamples}
\end{figure}

From the principal components we used the loadings ($L_0$, Eq.~\ref{L0}) and created the {\em sample} loading map (Figure~\ref{fig:PCASamples}).
In this loading map now each principal component is visualized in terms of a stacked histogram of the loadings of each sample, with the samples ordered according to their cell type. 
Furthermore, we also colored the samples with a common cell type. The visualization in Figure~\ref{fig:PCASamples} clearly shows that about 80\% of the loadings of
the second up to the fifth principal component that are attributed to the trisomic cell types, dominated by the trisomy 21 cell type. This result supports the finding of Altug-Teber {\em et al.} \cite{Altug2008}, who showed that the trisomic samples can be clearly distinguished from the control samples.

\section{Discussion}
We have presented two new visualization methods, loading maps and $l$-neighbor\-hood evolution matrices ($l$-NEMs), that assist linear and nonlinear feature extraction as well as an interactive framework, SpRay, that allows the visual exploration and comparison of many different dimensionality reduction methods. Our application to the study of the bacterium \emph{Streptomyces coelicolor} has confirmed that both visualization techniques are of great value for detection of important features and experiments in large and complex data matrices. Using these methods, we were able to identify time points that go along with major metabolic changes and to extract clusters of genes that are responsible for these changes. We have seen that loading maps provide an effective second perspective that complements the parallel coordinates, facilitates their interpretation and explains their appearance. Furthermore, we could show that the application of only one method is often not enough by demonstrating that PCA alone was unable to isolate a vital event after 36 hours (the change from primary metabolism to secondary metabolism as a result of phosphate depletion). In contrast, the use of neighborhood evolution matrices indicated the importance of this time point at first glance. A more detailed investigation that included the application of ICA with subsequent production of ICA-loading maps was then able to identify the genes  responsible for causing this event, namely phosphate-dependent genes and ribosomal genes. This demonstrates two things: First, that different methods are able to unveil different signals and, second, that the search for interesting signals is an interactive challenge/response procedure which makes it necessary to combine results, to rethink them in view of some new analysis and to see one result as a starting point for further steps.  

In the second application we applied our loading map visualization to a comparison of different cell types. While in the first application example the genes were the variables, in this case the samples are the variables. The loading map of the samples allow conclusions whether certain sample types can be distinguished from others based on their expression profiles. Furthermore, we demonstrated that the stacked histogram is not focused on spatial relations.

In this work, we have so far restricted our examples concerning the practical applicability of loading maps and $l$-neighborhood evolution matrices to linear feature extraction and the locally linear embedding algorithm, respectively. However, both techniques are useful in a much broader sense and we want to give some examples of further applications.

$l$-NEMs give an excellent overview of the mutual location of points in $n$-dimensional space which easily identifies clusters, connections and distributions. This is especially the case if the points are ordered. These insights are of great use in all algorithms that depend on the construction of neighborhood graphs. The latter, in turn, are an essential part of many manifold learning algorithms including LLE, Isopmap, LTSA and MVU in order to describe the local geometry around a point.

Loading maps can be generalized to other situations just as well. We already gave three more examples in the form of Eqs.~\ref{L1}-\ref{C1}. If we think of the measures as entries of a matrix, we see that this type of visualization is actually not limited to these four measures but works for every stochastic matrix. This includes all matrices that emerge from orthogonal matrices by squaring all of their entries. For the same reason, it is possible to combine our loading maps with other post-processing steps. For instance, the varimax criterion calculates an additional orthogonal transformation such that the sum of the variances of the loadings is maximized \cite{Kaiser1958}. As the product of two orthogonal matrices is an orthogonal matrix, the new loadings obviously still fulfill the requirements for a loading map and can thus easily be visualized in the same way.

Regarding scalability of our implementation, a closer look at the feature extraction methods presented here reveals that all of them rely at some point on an eigenvalue or singular value decomposition. It is well known that both decompositions are computationally demanding tasks \cite{Golub1983}. For biological data, however, we usually have data matrices $X\in \mathbb{R}^{n\times p}$ with $p\ll n$, a situation in which (under the agreement of a reduction of rows) the calculation of a thin SVD or the diagonalization of a $p \times p$ matrix is often feasible within a reasonable amount of time. For this reason, the computational complexity is in many relevant cases not an issue and our implementation scales well up to $2\cdot 10^4$ rows, with the columns ranging from $10-500$.

In some applications, it seems to be more appropriate to use slight variations of the visualization methods proposed. We therefore experimented with different modifications to the original loading maps and $l$-NEMs. For instance, it might sometimes be misleading that loadings belonging to a particular variable are, when viewed across several principal components, not necessarily at the same height. This is a consequence of the difference in the summed up heights of the preceding boxes of the stacked histogram. To eliminate this (not necessarily undesired) effect, one could instead consider representing the loadings through boxes of equal height, with the magnitude of the loadings visualized by a varying degree of transparency. This is, however, only feasible in the case where each box is visible, that is when the number of variables is not too large. When the variables represent all genes of a genome, as shown in our first application example, we consider our presented solution more appropriate.

Currently, $l$-NEMs differentiate between the neighbors of a particular data point through the use of a discrete set of colors. This is a plausible approach as long as the number of neighbors displayed is relatively small, say, not larger than 9-12. This is also the maximum number of colors that ColorBrewer offers for a specific palette. Otherwise, it is going to get harder and harder to find an adequate set of colors that can be easily distinguished by the eye. For larger sizes of the neighborhood, we propose using either a linear RGB gradient between two colors or, again, a transparency-based encoding.

In conclusion, feature extraction followed by dimension reduction offers a helpful overview of high-dimensional systems biology data, such as expression data. We have shown, however, that no universal solution with automatic interpretation exists, rather analysis of the data should be understood as an interactive process. This process includes statistical and visual methods. With the loadings maps and the LLE evolution matrices, we introduced two new visualization techniques, that are highly suitable for the analysis of neighborhoods and loadings in the biological context, and therefore they are a valuable complement to the linear and non-linear feature extraction methods.

\begin{acknowledgements}
We wish to thank Peter Wills for critically reading an earlier version of the manuscript. We also thank the anonymous referees for their thorough reviews and valuable comments that helped improving the presentation of the paper.
\end{acknowledgements}

%\pagebreak
%\bibliography{paper}   % name your BibTeX data base

%\begin{thebibliography}{}
% etc
%\end{thebibliography}

\end{document}